\documentclass[12pt, a4paper]{article}
\usepackage[latin1]{inputenc}
\usepackage[dvips]{graphicx,color}
\usepackage{amsmath}
\usepackage{amsfonts}
\usepackage{amssymb}
\usepackage{rotating}
\numberwithin{equation}{section}
\begin{document}
\def\pplogo{\vbox{\kern-\headheight\kern -29pt
\halign{##&##\hfil\cr&{\ppnumber}\cr\rule{0pt}{2.5ex}&\ppdate\cr}}}
\makeatletter
\def\ps@firstpage{\ps@empty \def\@oddhead{\hss\pplogo}%
  \let\@evenhead\@oddhead 
}
\def\maketitle{\par
 \begingroup
 \def\thefootnote{\fnsymbol{footnote}}
 \def\@makefnmark{\hbox{$^{\@thefnmark}$\hss}}
 \if@twocolumn
 \twocolumn[\@maketitle]
 \else \newpage
 \global\@topnum\z@ \@maketitle \fi\thispagestyle{firstpage}\@thanks
 \endgroup
 \setcounter{footnote}{0}
 \let\maketitle\relax
 \let\@maketitle\relax
 \gdef\@thanks{}\gdef\@author{}\gdef\@title{}\let\thanks\relax}
\makeatother
\def\eq#1{Eq. (\ref{eq:#1})}
\newcommand{\nc}{\newcommand}
\def\theequation{\thesection.\arabic{equation}}
\nc{\beq}{\begin{equation}}
\nc{\eeq}{\end{equation}}
\nc{\barray}{\begin{eqnarray}}
\nc{\earray}{\end{eqnarray}}
\nc{\barrayn}{\begin{eqnarray*}}
\nc{\earrayn}{\end{eqnarray*}}
\nc{\bcenter}{\begin{center}}
\nc{\ecenter}{\end{center}}
\nc{\ket}[1]{| #1 \rangle}
\nc{\bra}[1]{\langle #1 |}
\nc{\0}{\ket{0}}
\nc{\mc}{\mathcal}
\nc{\etal}{{\em et al}}
\nc{\GeV}{\mbox{GeV}}
\nc{\er}[1]{(\ref{eq:#1})}
\nc{\onehalf}{\frac{1}{2}}
\nc{\partialbar}{\bar{\partial}}
\nc{\psit}{\widetilde{\psi}}
\nc{\Tr}{\mbox{Tr}}
\nc{\tc}{\tilde c}
\nc{\tk}{\tilde K}
\nc{\tv}{\tilde V}
\nc{\CN}{{\mathcal N}}
\nc{\muphi}{\mu_{\phi}}

%

\setcounter{page}0
\def\ppnumber{\vbox{\baselineskip14pt
\hbox{\footnotesize{RUNHETC-2008-20, SLAC-PUB-13467}}
}}
\def\ppdate{\footnotesize{NSF-KITP-08-143, MIFP-08-28}} \date{}
\author{Rouven Essig$^{1}$, 
Jean-Fran\c{c}ois Fortin$^2$, Kuver Sinha$^{3}$, \\ Gonzalo Torroba$^{2,\,4}$, Matthew J.~Strassler$^2$\\
[7mm]
{\normalsize $^1$ SLAC National Accelerator Laboratory, Stanford University,}\\
{\normalsize Stanford, CA 94309, USA}\\
{\normalsize $^2$ NHETC and Department of Physics and Astronomy,}\\
{\normalsize Rutgers University, Piscataway, NJ 08855--0849, USA}\\
{\normalsize $^3$ Department of Physics, Texas A\&M University,}\\
{\normalsize College Station, TX 77843-4242, USA}\\
{\normalsize $^4$ Kavli Institute for Theoretical Physics, University of California,}\\
{\normalsize Santa Barbara, CA 93106--4030, USA}\\
}

\title{\bf Metastable supersymmetry breaking\\
[2mm]
and multitrace deformations of SQCD
\vskip 0.4cm}
\maketitle

\begin{abstract} \normalsize
\noindent

Metastable vacua in supersymmetric QCD in the presence of single and
multitrace deformations of the superpotential are explored, with the
aim of obtaining an acceptable phenomenology. The metastable vacua
appear at one loop, have a broken R-symmetry, and a magnetic gauge
group that is completely Higgsed.  With only a single trace
deformation, the adjoint fermions from the meson superfield are
approximately massless at one loop, even though they are massive at tree level and
R-symmetry is broken.  Consequently, if charged under the standard model,
they are unacceptably light.  A
multitrace quadratic deformation generates fermion masses proportional
to the deformation parameter.  Phenomenologically viable models of
direct gauge mediation can then be obtained, and some of their
features are discussed.
\end{abstract}
\bigskip
\newpage

\tableofcontents

\vskip 1cm

\section{Introduction} \label{sec:intro}

Relaxing the requirement that supersymmetry breaking occurs in the true vacuum (see e.g.~\cite{original}--\cite{Banks:2005df}) can help overcome many of the constraints of dynamical supersymmetry breaking with no supersymmetric vacua \cite{Thomas:1998jf}. Recently, Intriligator, Seiberg and Shih~\cite{iss} have shown that metastable dynamical supersymmetry breaking is rather generic and easy to achieve.
They found that metastable vacua occur in supersymmetric QCD (SQCD), in the free magnetic range, when the quarks have small masses,
\beq\label{eq:W1}
W= {\rm tr} \big(m \tilde Q Q\big)\,.
\eeq
This has opened many new avenues for model building and gauge mediation; see~\cite{Banks:2006ma}--\cite{Intriligator:2008fe} for some examples of recent work, and~\cite{Intriligator:2007cp} for a review and a more complete list of references.

It is not possible to build a phenomenologically viable model of gauge mediation using directly the ISS superpotential (\ref{eq:W1}).
This is due to an unbroken R-symmetry that forbids non-zero gaugino masses.
A natural question is then how the phenomenology changes when the superpotential is a more general polynomial in $\tilde Q Q$.
While this has been considered before for some particular superpotential deformations (see e.g.~\cite{kitano, terning, kutasov, giveon, zohar}), a more detailed account of the space of metastable vacua and the low energy phenomenology is needed. For instance, the light fermions of the model have not been fully explored. The aim of this work is to analyze the IR properties of the theory and its phenomenology in the presence of a generic $U(N_f)$-preserving
polynomial superpotential
\begin{equation}\label{eq:electricdefintro}
 W=m\,{\rm tr}(Q\tilde Q)+\frac{1}{2 \Lambda_0}\,{\rm tr}\,\big[(Q \tilde Q)^2\big]+\frac{1}{2 \Lambda_0}\,\gamma \big[{\rm tr}\,(Q \tilde Q)\big]^2 +\ldots\;\,,
\end{equation}
where $\Lambda_0 \gg \Lambda$ is some large UV scale, $\gamma$ is an order one coefficient, and `$\ldots$' are sextic and higher dimensional operators.

\vskip 2mm

Deforming (\ref{eq:W1}) by a generic polynomial in $\tilde Q Q$ breaks R-symmetry explicitly at tree level, and additional supersymmetric vacua are introduced~\cite{Nelson:1993nf}. The supersymmetric vacua for a single trace superpotential were analyzed in detail in~\cite{kutasov}, where it was found that the magnetic theory has classical supersymmetric vacua with various possible unbroken subgroups of the magnetic gauge group. This should be contrasted with the case of ISS, Eq.~(\ref{eq:W1}), where the magnetic gauge group is completely Higgsed and supersymmetry is broken classically by the rank condition.

After taking into account one loop quantum corrections in the magnetic theory, one finds the deformed theory also has metastable vacua at low energies~\cite{kutasov}. The dynamical reason for this is that the deformations to the magnetic superpotential come from irrelevant operators in the electric theory, which are parametrically suppressed. Therefore, we end up with a controllable deformation of the ISS construction in the IR. These vacua break R-symmetry spontaneously, and in phenomenologically interesting regions of parameter space the spontaneous breaking is much larger than the explicit breaking.

Since supersymmetric vacua allow for unbroken magnetic gauge groups, one might expect the same to occur for metastable vacua. However, the metastable vacua in the theories we explore below have a completely broken magnetic gauge group;
vacua with unbroken subgroups of the magnetic gauge group do not occur.
This is in some disagreement with~\cite{kutasov} and it would be interesting to see how this effect appears in the brane constructions of metastable vacua~\cite{Giveon:2007fk}.

Next we will analyze the phenomenological properties of the spectrum,
with particular attention to the light fermions, including the
Standard Model gauginos and a multiplet of fermions from the ``meson''
superfield $M=\tilde Q Q$.  If the superpotential contains only single
traces of powers of $M$, the singlet and adjoint parts of the meson
superfield $M=\tilde Q Q$ have the same one loop effective action.
The singlet fermion is the Goldstino, and must be massless at one loop
through a cancellation of its nonzero tree level mass against a one
loop correction.  The adjoint fermions (or more precisely, a certain
subset thereof) have the same tree and one loop effective action, and
so their masses arise only at two loops (and/or through equally small
mixing effects.)  Consequently their masses are small compared with
those of the Standard Model gauginos, which arise at one loop.

In this paper we will be considering the case where the embedding of
the Standard Model gauge group into the $U(N_f)$ flavor group endows
these fermions with Standard Model quantum numbers.  With such light
masses, these fermions would already have been observed, and so these
models would be phenomenologically unacceptable.

We are therefore led to consider a multitrace deformation of the
superpotential; in particular, we must take $\gamma\neq 0$ in
Eq.~(\ref{eq:electricdefintro}).  Then the cancellation between the
tree level and one loop masses for the Goldstino fails for the adjoint
fermions, leaving them with masses proportional to $\gamma$.  The
phenomenology of direct gauge-mediated models based on this theory is
quite rich, since the adjoint fermions
may be lighter or heavier than the
Standard Model gauginos, depending on $\gamma$.  Mixing between these
fermions and the gauginos is negligibly tiny, due to a
charge-conjugation symmetry in (\ref{eq:electricdefintro}).  We will
briefly discuss some of the interesting phenomenological properties of
such a scenario, leaving the details to a forthcoming publication
\cite{EFSST2}.

The various sections are arranged as follows.  In Section \ref{sec:sqcd}, we
discuss the moduli space of SQCD with the superpotential
Eq.~(\ref{eq:electricdefintro}), keeping only terms up to quartic order in
the electric fields.  In Section \ref{sec:iss}, we review SQCD without
deformations (ISS), with emphasis on the spectrum and associated
phenomenological issues.  In Section \ref{sec:single}, we study single trace
deformations of the ISS superpotential, that is, the case
$\gamma=0$. We show that all metastable vacua have a magnetic gauge
group that is completely Higgsed, and we discuss the spectrum, showing it
is unacceptable for phenomenology.  Next, in Section \ref{sec:multi} we consider
$\gamma\neq 0$, describing the spectrum in detail.  Finally, Section \ref{sec:pheno}
contains a brief overview of the phenomenology of and constraints on
such models.  Various computations are shown in detail in the
Appendix.

\section{SQCD with a multitrace superpotential} \label{sec:sqcd}

In this section, we analyze the symmetries and supersymmetric vacua of SQCD in the presence of a generic $U(N_f)$-preserving polynomial superpotential.

Supersymmetric QCD with gauge group $SU(N_c)$ and $N_f$ flavors $(Q_i, \tilde Q_j)$ with equal masses $m$ has a global symmetry group
\beq
SU(N_f)_V \times U(1)_V
\eeq
under which $(Q_i,\,\tilde Q_i)$ transform as $(\Box_{+1},\,\overline \Box_{-1})$. There is also a discrete $\mathbb Z_2$ charge conjugation symmetry $Q_i \leftrightarrow \tilde Q_i$.  For phenomenological applications we will later weakly gauge a subgroup of $SU(N_f)_V$ and identify it with the Standard Model gauge groups.  We will also gauge $U(1)_V$ to remove a Nambu-Goldstone boson.

The most general quartic superpotential preserving this symmetry is
of the form
\begin{equation}\label{eq:electricdef}
 W=m\,{\rm tr}(Q\tilde Q)+\frac{1}{2 \Lambda_0}\,{\rm tr}\,\Big[(Q \tilde Q)^2\Big]+\frac{1}{2 \Lambda_0}\,\gamma \Big[{\rm tr}\,(Q \tilde Q)\Big]^2 \,.
\end{equation}
We will typically consider $\Lambda_0 \gg \Lambda \gg m$, and take
$\gamma$ to be of order one or smaller.  We will not consider sextic
or higher operators, since they are suppressed by higher powers of
$\Lambda_0$ and would not affect our discussion. The
nonrenormalizable superpotential (\ref{eq:electricdef}) could be
generated from a renormalizable theory, for example by integrating out
fields with masses  $\sim\Lambda_0$ that couple to $Q \tilde
Q$.

Let us consider the theory in various limits.  First, for $W=0$ there
is a moduli space of vacua parameterized by mesons and baryons modulo
classical constraints.  The global symmetry is enhanced to
$SU(N_f)_L\times SU(N_f)_R\times U(1)_V$, and there is a non-anomalous
$U(1)_R$ symmetry as well as an anomalous $U(1)_A$ axial current.

For $m/\Lambda \neq 0$ but $\Lambda_0 \to \infty$, the superpotential is
renormalizable, and the theory has an exact classical $U(1)_R$ symmetry which
is anomalous
at the quantum level.\footnote{There is also an approximate non-anomalous
R-symmetry ``$U(1)_{R'}$'' which is restored as $m\to 0$, but we will not need to
consider
this symmetry.} The non-anomalous symmetries of the model are
\begin{center}
\begin{tabular}{c|ccc}
&$SU(N_f)_V$&$U(1)_R$&$U(1)_V$\\
\hline
$Q_i$&$\Box$&$+1$&$+1$\\
$\tilde Q_i$&$\overline \Box$&$+1$&$-1$\\
$\Lambda^{3N_c-N_f}$&$0$&$2N_c$&$0$
\end{tabular}
\end{center}
plus the $\mathbb Z_2$ charge conjugation. The F-term relations lift the moduli space and the only vacuum is at the origin.

On the other hand, for $m\neq0$ and $\Lambda_0$ large but finite, all R-symmetries are explicitly broken at the classical level.
New discrete supersymmetric vacua appear in the regime
$$
\tilde Q \, Q \sim m \Lambda_0\,.
$$

\subsection{Magnetic dual}\label{subsec:magnetic}

Below the scale $\Lambda$, the theory is described by an effective
theory, called the ``dual magnetic theory'', with gauge group
$SU(\tilde N_c)$,
singlet mesons $\Phi_{ij}$, and $N_f$ fundamental
flavors $(q_i, \tilde q_j)$; we define $\tilde N_c\equiv N_f-N_c$.
The theory has a positive beta function
and is weakly-coupled in the infrared.  After an appropriate change of
variables, the classical tree level superpotential reads
\begin{equation}\label{eq:magneticdef}
W =
h \,{\rm tr}(q \Phi \tilde q)-h \mu^2\,{\rm tr}\,\Phi+
+\frac{1}{2}h^2 \mu_\phi \Big({\rm tr}\,\Phi^2+\gamma({\rm tr}\,\Phi)^2 \Big)\,.
\end{equation}
where the first trace is over magnetic color and the remaining
traces are over flavor indices.
The relation with the electric variables is (roughly)
$$
\Lambda \Phi \sim \tilde Q Q,\;h\,\mu^2 \sim \Lambda m\;,\;h^2\,\mu_\phi \sim \frac{\Lambda^2}{\Lambda_0}\,.
$$
More details may be found in~\cite{iss}.

As in ISS, we restrict to small quark masses $m \ll \Lambda$. We will also restrict ourselves to the range
\begin{equation}\label{eq:cond1}
\Lambda_0 \gg \sqrt{\frac{\Lambda}{m}}\,\Lambda\,,
\end{equation}
which guarantees that $h \mu_\phi \ll \mu$. This will be needed to have long-lived metastable vacua. There are nonperturbative corrections to the superpotential (\ref{eq:magneticdef}), but they are all small enough not to affect our calculations given (\ref{eq:cond1}).

Also, these conditions ensure that the symmetries of the model at the
scale $\Lambda$ are approximately $SU(N_f)_L\times SU(N_f)_R\times
U(1)_V \times U(1)_{R'}$, broken to $SU(N_f)_V\times U(1)_V$ only by
effects of order $m/\Lambda$ and $\Lambda/\Lambda_0$.  Therefore,
to an excellent approximation,
both the superpotential {\it and} the K\"ahler potential
satisfy the larger symmetry group, under which the trace and traceless
parts of $\Phi_{ij}$ transform as a single irreducible multiplet.  We will
work only to leading non-vanishing order in the symmetry-breaking
effects from non-zero $m$ and non-infinite $\Lambda_0$.

Furthermore, the discrete $\mathbb Z_2$ charge-conjugation symmetry of the
electric theory appears as the transformation
\begin{equation}\label{eq:Csym}
 \Phi \to \Phi^T\,,\;q_i \leftrightarrow \tilde q_i \ .
\end{equation}
This transformation plays an important role in
the phenomenology of gauge mediation models
based on (\ref{eq:magneticdef}), and indeed in other
ISS-related models (see e.g.~\cite{Fox:2002bu}).

As in the electric theory, the R-symmetry is explicitly broken, and we expect new supersymmetric vacua parametrically at $\mu^2/\mu_\phi$. Indeed, the solutions to the F-term constraints
\begin{eqnarray}\label{eq:F-term1}
\big(-h \mu^2+ h^2 \mu_\phi \gamma\,{\rm tr}\,\Phi \big)\,I_{N_f \times N_f}+ h^2 \mu_\phi\,\Phi+h\,\tilde q q&=&0\nonumber\\
q \Phi= \Phi \tilde q&=& 0\,,
\end{eqnarray}
are
\begin{equation}\label{eq:susyPhi}
\langle h\,\Phi \rangle = \frac{1}{1+(N_f-k)\gamma} \, \frac{\mu^2}{\muphi} \left(\begin{matrix}  0_{k \times k} & 0_{k \times (N_f - k)} \\ 0_{(N_f - k) \times k} & I_{(N_f - k)\times(N_f - k)} \end{matrix}\right)
\end{equation}
and
\begin{equation}\label{eq:susyq}
\langle \tilde q q \rangle = \frac{1}{1+(N_f-k)\gamma}\, \mu^2\, \left(\begin{matrix} I_{k\times k} & 0_{k \times (N_f - k)} \\ 0_{(N_f - k) \times k} & 0_{(N_f - k) \times (N_f - k)}  \end{matrix}\right)
\end{equation}
with $k=1,\ldots, N_f- N_c$.  (Here $I$ represents the identity matrix, and
a subscript $r\times s$ indicates a block matrix of the corresponding size.)
The appearance of the extra parameter $k$ classifying different classical vacua has been observed for $\gamma=0$ by~\cite{kutasov}.  In particular, for $k< N_f- N_c$ there is an unbroken magnetic gauge group $SU(N_f -N_c -k)$.

\section{Metastable DSB in the R-symmetric limit} \label{sec:iss}

In the next three sections, we will analyze the IR dynamics of
(\ref{eq:magneticdef}) in three steps.  First, we review the ISS model
\cite{iss}, the R-symmetric limit $\mu_\phi=0$, which corresponds to
an electric SQCD with massive flavors and no irrelevant operators. We
will highlight the spectrum and associated phenomenological
problems. In Section \ref{sec:single}, we show how these problems are
not entirely solved by making $\mu_\phi$ non-zero but leaving $\gamma=0$.
Finally, in Section \ref{sec:multi}, we show how the theory with
$\gamma\neq 0$ resolves the remaining problems.

\subsection{The model and its spectrum} \label{subsec:review}

The ISS model considers massive SQCD near the origin in field space in the free magnetic range $N_c+1 \le N_f < \frac{3}{2} N_c$, where the theory has a dual magnetic description with superpotential
\begin{equation}
W  = -h \mu^2\,{\rm tr}\,\Phi+ h {\rm tr}(q \Phi \tilde q)\,.
\end{equation}
At the classical level the theory breaks supersymmetry by the rank
condition.  We parametrize the fields by
\begin{equation}\label{eq:paramPhi}
\Phi= \left(\begin{matrix} Y_{\tilde N_c \times \tilde N_c} & Z^T_{\tilde N_c \times N_c} \\ \tilde Z_{N_c \times \tilde N_c} &X_{N_c \times N_c}\end{matrix} \right)
\end{equation}
\begin{equation}\label{eq:paramq}
q^T=\left( \begin{matrix} \chi_{\tilde N_c \times \tilde N_c} \\ \rho_{N_c \times \tilde N_c} \end{matrix}\right)\;,\;\tilde q=\left( \begin{matrix} \tilde \chi_{\tilde N_c \times \tilde N_c} \\ \tilde \rho_{N_c \times \tilde N_c} \end{matrix}\right)\,,
\end{equation}
where $\tilde N_c= N_f- N_c$ is the rank of the magnetic gauge group.  The classical moduli space of vacua is parametrized by $\langle \chi \tilde \chi \rangle=\mu^2\,I_{\tilde N_c \times \tilde N_c}$ and $\langle X \rangle$. The other fields have vanishing expectation values. In the rest of the paper we will restrict to metastable vacua with maximal unbroken global symmetry, by choosing the ansatz
\begin{equation}\label{eq:X0}
\langle X\rangle= X_0 \,I_{N_c \times N_c}\;,\;
\langle \chi \rangle=q_0\,I_{\tilde N_c \times \tilde N_c}\;,\;
\langle \tilde \chi \rangle=\tilde q_0\,I_{\tilde N_c \times \tilde N_c}\,.
\end{equation}
It will be checked that this is a self-consistent choice.

The vev for $\chi \tilde \chi$ breaks the gauge group $SU(\tilde N_c)_G$ completely,
with the breaking pattern
\beq\label{eq:symm}
SU(\tilde N_c)_G \times SU(N_f)_V \times U(1)_V \to SU(\tilde N_c)_V \times SU(N_c) \times U(1)'\,.
\eeq
(Here all groups except $SU(\tilde N_c)_G$ are global; we remind
the reader that $\tilde N_c=N_f-N_c$).  The reduction of the global
symmetry group leads to
$2N_c
\tilde N_c+1$ Nambu-Goldstone modes. The fields $(\rho, \tilde \rho, Z, \tilde Z)$ are charged under $U(1)'$, which plays the role of a messenger number symmetry. See~\cite{iss} for a more detailed discussion.

The flat directions $X$ are not protected by holomorphy or symmetries and,
as we shall review shortly, become massive at one loop.  (A field
with these properties is called a ``pseudo-modulus''~\cite{iss}.)  In
particular, $X$ is stablized at the origin.  Near
the origin of moduli space the rank condition imposes
\begin{equation}\label{eq:rank}
|F_X|=|h \mu^2|\,,
\end{equation}
and the scale of supersymmetry breaking is
\begin{equation}
V_{min}=N_c\,|h^2 \mu^4|\,.
\end{equation}

To analyze the spectrum of the theory, it is convenient to rewrite the superpotential in terms of the component fields,
\begin{eqnarray}\label{eq:Wiss2}
W&=&-h\mu^2\,{\rm tr}\, X+h\,{\rm tr}\;\left( \begin{matrix} \rho& Z\end{matrix} \right) \left( \begin{matrix}  X& \mu \\ \mu & 0\end{matrix} \right) \left( \begin{matrix} \tilde \rho\\ \tilde Z\end{matrix} \right)\nonumber\\
&&+\,h\mu\,{\rm tr}\big[Y(\chi+\tilde \chi) \big]+h\,{\rm tr}\big(\chi Y \tilde \chi+ \rho \tilde Z \tilde \chi+\chi Z \tilde \rho \big)\,.
\end{eqnarray}
The spectrum consists of three sectors, each consisting of
fields satisfying ${\rm Str}\,M^2 = 0$.

(1) \emph{The $(\rho, Z)$ sector}: Treating $X$ as a background superfield, the $(\rho, Z)$ supersymmetric mass matrix is
\begin{equation}\label{eq:fmassmatrix}
M_f=\left( \begin{matrix} h X& h \mu \\ h\mu & 0 \end{matrix} \right)
\end{equation}
while the bosonic matrix is computed, as usual, including off-diagonal blocks with F-terms.

There are $2N_c \tilde N_c$ Dirac fermions that come from $(\psi_\rho,\,\psi_Z)$ and $(\psi_{\tilde \rho},\,\psi_{\tilde Z})$.  Near the origin of field space, their masses are of order $h\mu$, from (\ref{eq:fmassmatrix}). The scalars combine into $4 N_c \tilde N_c$ complex fields, which are linear combinations of $(\rho, Z, \tilde \rho^*, \tilde Z^*)$. There are $N_c \tilde N_c$ complex Nambu-Goldstone bosons from the combinations ${\rm Re}\,(\rho+\tilde \rho)$ and ${\rm Im}\,(\rho-\tilde \rho)$. The $3 N_c \tilde N_c$ remaining complex scalars have splittings of order, and centered around, $h \mu$. The numerical coefficients adjust to preserve ${\rm Str}\,M^2=0$.

This sector will play the role of the messenger sector in gauge
mediation applications. Once a subgroup of the flavor symmetry is
identified with the Standard Model, and gauged with couplings
$g_{SM}$, the Nambu-Goldstone modes will acquire a one loop mass of
order $g_{SM} \mu/(4\pi)$.  (In particular, we will study the
case where $SU(N_c)$ is gauged --- see Eq.(\ref{eq:symm}).) The lightest state will be stable in
the full theory, since the messenger sector is protected by the
non-anomalous $U(1)'$ messenger number.

(2) \emph{The $(Y,\chi)$ sector}: Fermions from $Y, (\chi+\tilde \chi)$ form $\tilde N_c^2$ Dirac fermions with mass $\sim h\mu$. The traceless part\footnote{We denote traceless fields with primes; for instance $X'$ is the traceless part of $X$.} of the chiral superfield $(\chi-\tilde \chi)$, which contains the NG bosons ${\rm Im}\,(\chi'-\tilde \chi')$, is eaten by the superHiggs mechanism when the magnetic group is gauged.

The field ${\rm Im}\,{\rm tr}(\chi - \tilde \chi)$ is a NG boson associated to the breaking of $U(1)_V$. The field ${\rm Re}\,{\rm tr}(\chi - \tilde \chi)$ corresponds to a pseudo-modulus, which will be lifted at one loop. The fermion from ${\rm tr}\,(\chi -\tilde \chi)$ is massless. This sector has a supersymmetric spectrum at tree level.

The massless fields from ${\rm tr}\,(\chi-\tilde \chi)$ would be
phenomenologically forbidden. This forces us to gauge $U(1)_V$, so
that the superfield ${\rm tr}\,(\chi-\tilde \chi)$ is eaten by the
$U(1)_V$ gauge boson and at tree level acquires a mass of order $g_V
\mu$.

(3) \emph{The $X$ sector}: $X$ is a flat direction, with massless fermionic partner at tree level. In particular, $\psi_{{\rm tr}\,X}$ is the Goldstino.

\vskip 2mm

One loop contributions from heavy particles
lift the pseudo-moduli.  The fields $(Y, \chi, \tilde \chi)$ do not couple at tree level to the supersymmetry breaking sector, so they do not contribute to the one loop effective potential for the pseudo-moduli. Because we are in the regime where $|F_X|=|h \mu^2|$ is of order the square of the messenger masses, the effect of integrating out the messengers does not have a simple expression in superspace, and it is more convenient to work directly with nonsupersymmetric expressions. The bosonic action is given by the usual Coleman-Weinberg formula~\cite{cw}
\begin{equation}\label{eq:Vcw1}
V_{CW}=\frac{1}{64 \pi^2}\,{\rm STr}\,M^4\,{\rm log}\,\frac{M^2}{\Lambda^2}\,.
\end{equation}

Near the origin of moduli space $X \ll \mu$, the potential is approximated by~\cite{iss}
\begin{equation}\label{eq:Vcw2}
V_{CW} \approx \frac{a}{2}\,|h^4 \mu^2|\,{\rm tr}\,\left( {\rm Re}\,\frac{1}{\sqrt{2}}[\chi - \tilde \chi]\right)^2+ b |h^4 \mu^2|\,{\rm tr}\,(X^\dag X)
\end{equation}
with
\beq\label{eq:b}
a=\frac{{\rm log}\,4-1}{8 \pi^2}\, N_c\;,\;\; b=\frac{{\rm log}\,4-1}{8 \pi^2}\, \tilde N_c\,.
\eeq
Therefore, in the ISS model the pseudo-moduli are consistently stabilized at the origin and R-symmetry is preserved. In this approximation, the one loop mass of the bosonic field $X$ is given by
\begin{equation}\label{eq:mcw}
m_{CW}^2= b |h^4\mu^2|
= \frac{{\rm log}\,4-1}{8 \pi^2}\, \tilde N_c\,|h^4 \mu^2|\,.
\end{equation}

\subsection{Phenomenological problems} \label{subsec:spectrum2}

One could try to use the ISS construction as the supersymmetry breaking sector in models of direct gauge mediation. However, since R-symmetry is preserved in the metastable vacuum, Majorana masses for the Standard Model gauginos are forbidden. The same applies to the fermions $\psi_X$ and $\psi_{\chi -\tilde \chi}$, which may have SM quantum numbers after embedding the SM gauge group into the flavor symmetry group of the model. For these reasons, this model does not give an acceptable phenomenology.

There are various ways of improving this situation (see, for instance,~\cite{Abel:2007jx}--\cite{Marques:2008va}). One very interesting proposal~\cite{Fox:2002bu} is that the gauginos could come from Dirac fermions, whose mass is not constrained to vanish by an unbroken R-symmetry. This idea was applied to the ISS model in~\cite{Amigo:2008rc}, by adding new fields and interactions to the superpotential. Dirac masses appear from one loop diagrams mixing the MSSM Weyl gauginos with the new Weyl fermions. One problem with this approach is that doubling the number of fields (in order to have Dirac fermions) creates a Landau pole close to the messenger scale. In this case, corrections from the microscopic theory may become important.

Another possibility is to deform the superpotential by higher powers of the meson superfield, explicitly breaking the R-symmetry at tree level~\cite{kitano,
terning, kutasov}.  We consider this possibility in detail below.


\section{Single trace deformation} \label{sec:single}

We begin by considering  the superpotential Eq.~(\ref{eq:magneticdef})
with $\gamma=0$, that is, with only a single trace perturbation:
\begin{equation}\label{eq:single trace}
W=-h \mu^2\,{\rm tr}\,\Phi+ h {\rm tr}(q \Phi \tilde q)+\frac{1}{2}h^2 \mu_\phi\, {\rm tr}\,(\Phi^2)\,.
\end{equation}
This model was discussed in~\cite{kutasov}, where it was suggested that new metastable vacua, with unbroken magnetic group, appear around $X \sim \mu$. However, this region of parameter space is subtle, because higher order corrections to (\ref{eq:Vcw2}) become important.
We will have two new things to say about this model.

(1) By considering the full logarithmic one loop potential
    (\ref{eq:Vcw1}), it is possible to show that the metastable vacua
    with unbroken magnetic gauge group are actually unstable. Thus,
    one is led to study only the ISS-like vacuum where the magnetic
    gauge group is completely Higgsed.

(2) Gauginos indeed become massive at one loop in this model,
as expected from the R-symmetry breaking.  However (ignoring some
subtleties which we will discuss later) the adjoint fermions
$\psi_{X'}$ become massive only at two loops, because
diagrammatic cancellations that make the Goldstino $\psi_{{\rm tr}\,X}$
massless at one loop also force the adjoint fermions $\psi_{X'}$ to
be massless at this order.  This provides the main motivation for
studying non-zero $\gamma$ below.

\subsection{Metastable supersymmetry breaking}\label{subsec:k-cases}

The classical supersymmetric vacua are obtained by setting $\gamma=0$ in (\ref{eq:susyPhi}) and (\ref{eq:susyq}).
In order to analyze the effect of the deformation on the ISS metastable vacuum, the cases $k=N_f-N_c$ and
$k< N_f- N_c$ have to be distinguished.

\subsubsection*{Case $k=N_f-N_c$}

This is the analog of the ISS construction, with no unbroken gauge group. The fields are parameterized as in Eqs.~(\ref{eq:paramPhi}) and (\ref{eq:paramq}). We will now review why a metastable vacuum appears at a distance of order $\mu_\phi/b$ away from the origin~\cite{kutasov}.

\begin{figure}[t]
\begin{center}
\includegraphics[scale=0.83]{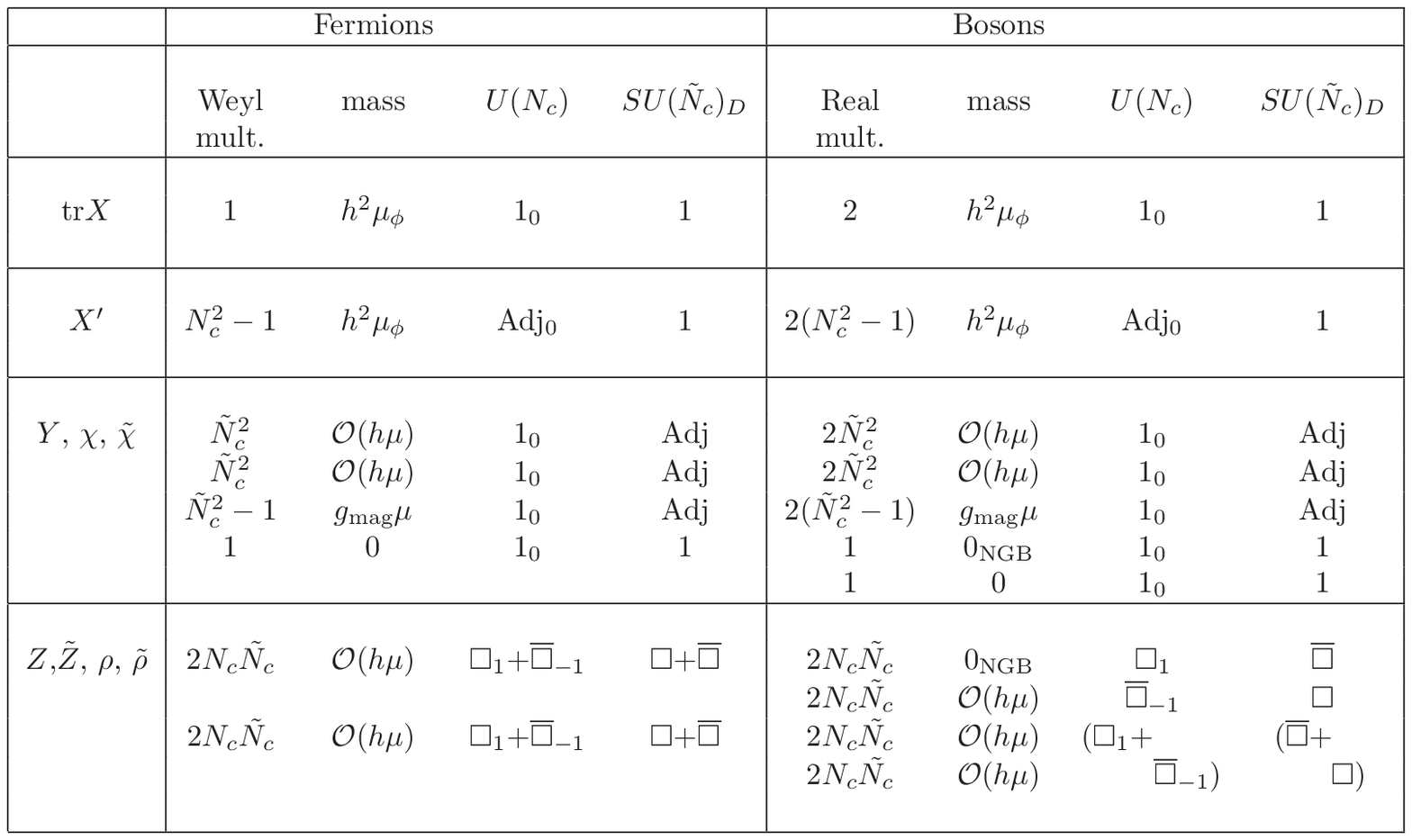}
\caption{The classical mass spectrum, grouped in sectors with ${\rm
Str} \,M^2=0$.  Since supersymmetry is spontaneously broken only after
including one loop effects, there is no Goldstino at tree level.  $g_{mag}$ is the magnetic gauge coupling.
A subscript ``NGB'' indicates the particle is massless because it is
a Nambu-Goldstone boson. Subscripts in the third column indicate the charge under the $U(1)$
subgroup.  Note this table gives the spectrum
before the Standard Model gauge group is gauged. }
\label{table:classicalmasses}
\end{center}
\end{figure}

As a starting point, set $V_{CW} \to 0$. Due to the classical
deformation, $X$ is no longer a flat direction, unlike the ISS
case. Rather, the origin $X_0 \sim 0$ is at the side of a paraboloid of
classical curvature $|h^2 \mu_\phi|^2$. In other words, the origin is unstable
against classical flow of $X_0$ toward
the supersymmetric
vacua discussed before. The tree level spectrum near the origin is
shown in Figure \ref{table:classicalmasses}.

In order to create a local minimum, the quantum contribution
$V_{CW}\sim m_{CW}\,|X_0|^2$ should overwhelm the curvature
of the classical potential,
i.e., $m_{CW} \gg |h^2\,\mu_\phi|$. This rather interesting effect,
where a one loop contribution stabilizes a classical runaway
direction, was analyzed in~\cite{est}. Here, the stabilization
of $X_0$ can occur naturally, since
$\mu_\phi$, arising  from a nonrenormalizable
operator in the microscopic theory, is
parametrically small. The condition that the one loop
potential introduces a supersymmetry breaking minimum,
\begin{equation}
 \epsilon \equiv \frac{m_{cl}^2}{m_{CW}^2} \approx \Big|\frac{\muphi^2}{b\mu^2}\Big| \ll 1\,,
\end{equation}
is naturally satisfied.

\begin{figure}[t]
\begin{center}
\includegraphics[scale=0.82,angle=0]{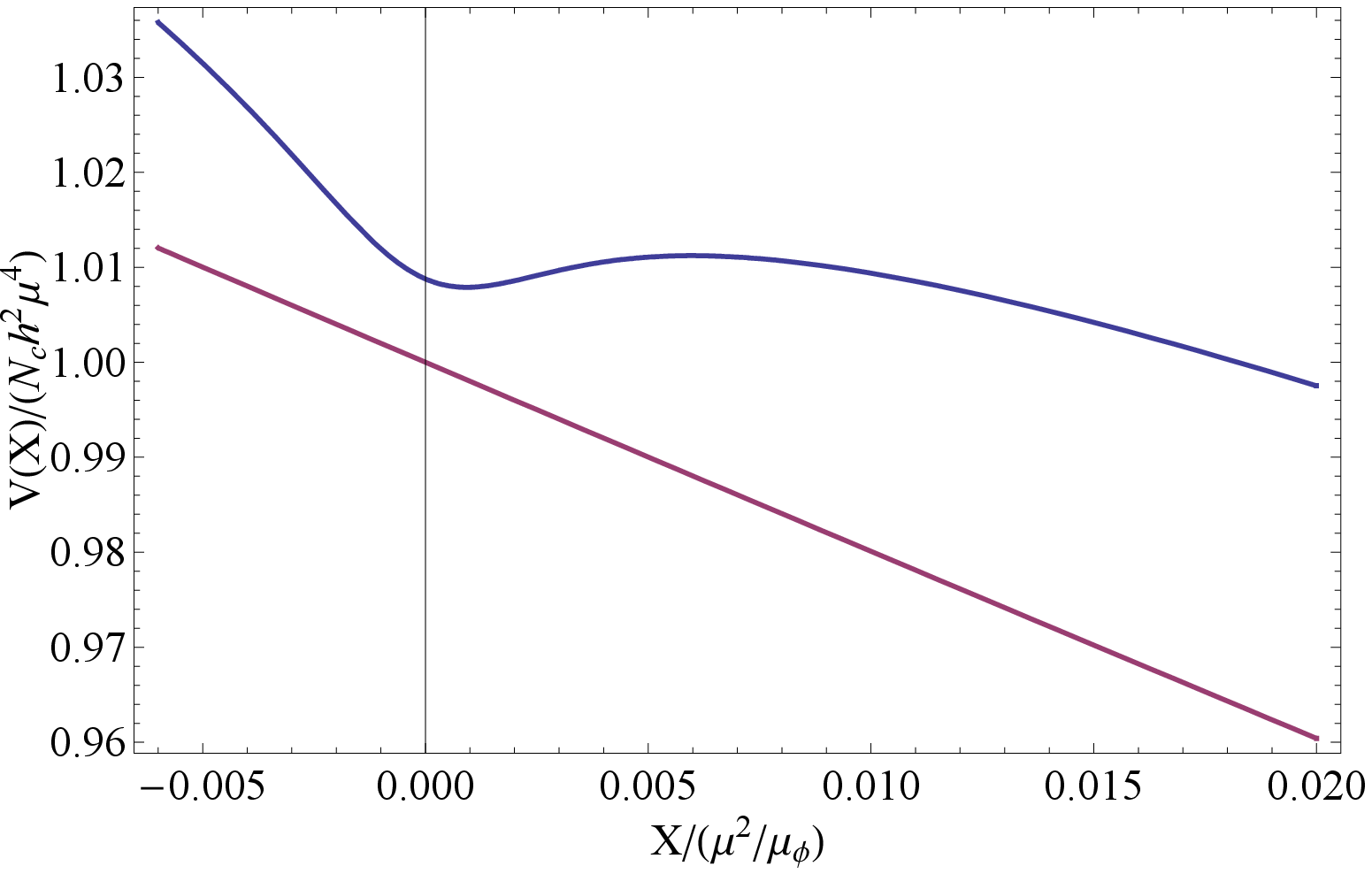}
\caption{Metastable vacuum near $X\sim 0$, for a single trace quadratic deformation
of the superpotential (i.e.~$\gamma=0$).  All parameters have been chosen to
be real.  The bottom (magenta) line is the tree level potential, while the
top (blue) line shows the tree level potential plus one loop Coleman-Weinberg
corrections.  The $X$-axis has been normalized such that the position of the
tree level supersymmetric vacuum lies at $X/(\mu^2/\muphi) = 1$.
Notice how the one loop corrections create a (metastable) minimum near the origin.
}
\label{fig:Vcw-slice}
\end{center}
\end{figure}

The potentials at tree level and at one loop, as a function of $X_0$,
are shown in Figure \ref{fig:Vcw-slice}.
As seen from the figure, the tree level potential (lower magenta curve), which is obtained from the superpotential in (\ref{eq:single trace}),
has no supersymmetry breaking minimum.
A metastable minimum is created near the origin once the one loop quantum corrections in the form of $V_{CW}$ are
included (upper blue curve).

As a result of the
competition between the classical and quantum contributions, a
metastable vacuum is created at
\begin{equation}\label{eq:metavac}
 h X_0  \approx \frac{\mu^2 \mu_\phi^*}{b|\mu|^2+|\mu_\phi|^2}\;,\;q_0 \tilde q_0=\mu^2\, ;
\end{equation}
see Eq.~(\ref{eq:X0}) for the notation.
As expected, $X_0$ is proportional to the explicit R-symmetry breaking parameter $\mu_\phi$. However, it is larger than this by the inverse loop factor $1/b$. This follows from the fact that the minimum appears from balancing a tree level linear term of order $\mu^2\,\mu_\phi$ against a one loop quadratic term of order $b \mu^2$.

The pattern of symmetry breaking in this vacuum is
\begin{equation}\label{eq:breaking}
SU(\tilde N_c)_G \times SU(N_f)_V \times U(1)_V \to SU(\tilde N_c)_V \times SU(N_c) \times U(1)'\,,
\end{equation}
where only the messengers transform under $U(1)'$. Unlike the ISS
construction, here $X_0 \neq 0$, so that the R-symmetry is both
explicitly and spontaneously broken, {\it with the latter dominating} since
$|h X_0| \gg |\mu_\phi|$.

\subsubsection*{Case $k<N_f-N_c$}

The possibility of metastable vacua with $k<N_f-N_c$ is very interesting; coupling this to the MSSM, it would imply unbroken gauge groups in the hidden sector. Properties of such configurations were discussed in~\cite{kutasov}. Unfortunately, we will now show that there are generically no metastable vacua in this regime.

Such vacua should be of the form
\begin{equation}\label{eq:kutmin}
\Phi= \left(\begin{matrix} 0 & 0 \\ 0 &X_{(N_f-k)\times(N_f-k)}\end{matrix} \right)\;,\;\tilde q q=\left(\begin{matrix} \mu^2 I_{k\times k} & 0 \\ 0 &0\end{matrix} \right)\,.
\end{equation}
The parametrization of the fluctuations is slightly more involved,
\begin{equation}\label{eq:param}
 \Phi=\left(\begin{matrix} Y_{k \times k}&Z_{k \times (N_f-k)}\\ \tilde Z_{(N_f-k) \times k}&X_{(N_f-k)\times (N_f-k)} \end{matrix}\right)\;,\;
 q=\left(\begin{matrix} V_{k \times k}& T_{k \times (\tilde{N}_c-k)}\\ P_{(N_f-k)\times k}&\varphi_{(N_f-k) \times (\tilde{N}_c-k)} \end{matrix}\right) \;
\end{equation}
and similarly for $\tilde q$. As in the case $k=N_f-N_c$, the expectation values are chosen to be of the form
$$
\langle X \rangle= X_0\, I_{(N_f-k)\times(N_f-k)}\;,\;\langle V \rangle= q_0 \,I_{k \times k}\;,\;\langle \tilde V \rangle= \tilde q_0 \,I_{k \times k}\,.
$$
The new fields $(\varphi, \, \tilde \varphi)$ and $(T, \,\tilde T)$ do not exist for $k=N_f-N_c$. They are fundamental flavors of the unbroken magnetic group $SU(N_f-N_c-k)$.

As was found in~\cite{kutasov}, positivity of the bosonic mass matrix of $(\varphi, \tilde \varphi)$ implies
$$
|X_0|^2 \ge |\mu^2-h \mu_\phi X_0|\,.
$$
This places us in the regime $X_0 \gtrsim \mu$. In this regime, the quadratic approximation (\ref{eq:Vcw2}) to the Coleman-Weinberg potential is no longer valid. For $X_0/\mu \sim 1$, all the higher order terms in $V_{CW}$ give contributions comparable to (\ref{eq:Vcw2}). In other words, it is necessary to use the full expression appearing in Eq.~(\ref{eq:Vcw1}).

Therefore, to establish the existence of such vacua, a detailed
analysis of $V_{CW}$ is required. As shown in the Appendix, all
such vacua are unstable once the full form of $V_{CW}$ is
included.   The intuitive reason for
this is that at large $X_0$ the logarithmic growth of $V_{CW}$ cannot
overwhelm the quadratic terms in
the classical potential. A similar behavior was found in~\cite{est}.

The plot of $V_{tree}+V_{CW}$ for this case is almost the same as that
of Figure \ref{fig:Vcw-slice}. For sufficiently large $|X_0/\mu| >1$,
the classical falling potential dominates the logarithmic rise of
the $V_{CW}$, and no critical points are found until the
supersymmetric vacuum is reached.

\vskip 2mm

Summarizing, metastable states occur only for $k=N_f-N_c$.
The fields have expectation values Eq.~(\ref{eq:metavac}),
breaking the magnetic gauge group completely at the
scale $\mu$.

\subsection{Light fermions}\label{subsec:lightf}

We therefore return to the one remaining vacuum, the ISS-like case
with $k=N_f-N_c$.
From the previous analysis, the bosons from $X$
and the traceless part of $\chi-\tilde \chi$ acquire masses of order
$m_{CW}$.  The aim of this section is to
compute the fermion masses at one loop, and show that
$\psi_{X_{ij}}$ remains massless at this order, contrary to
naive expectations from R-symmetry breaking.

First we explore one loop effects involving the Goldstino $\psi_{{\rm
tr}\,X}$. At tree level it has a nonvanishing mass $h^2
\mu_\phi$. We are not expanding
around a critical point of the classical potential, but rather one of
the full one loop potential, and therefore the Goldstino
should become massless only once one loop effects are included. This
implies that the one loop diagram has to give
\begin{equation}\label{eq:oneloop}
m^{1-loop}_{\psi_{{\rm tr}\,X}}\,\approx\,-h^2 \mu_\phi\,,
\end{equation}
such that $m^{tree}_{\psi_{{\rm tr}\,X}}+m^{1-loop}_{\psi_{{\rm tr}\,X}}\approx 0$. Indeed, the explicit evaluation of the one loop diagram in the Appendix corroborates (\ref{eq:oneloop}). These results are approximate because we are neglecting (subleading) mixings with other singlet fermions; see below and the Appendix.

At a first glance it is surprising that the one loop contribution can be equal to the tree level one. This is so because the one loop diagram is of order
$$
\frac{h^2}{16\pi^2}\, h \,X_0\,.
$$
However, since $h X_0 \sim \mu_\phi/b$, with $b$ defined
in Eq.~(\ref{eq:b}), we obtain the result (\ref{eq:oneloop}). This is another manifestation of the pseudo-runaway behavior discussed in the previous section.

Next, notice that within the classical superpotential (\ref{eq:single
trace}), $X_{ij}$ only appears in single traces. On the other hand,
the one loop contribution is a single trace of a function of $X_{ij}$,
because it comes from exponentiating bosonic and fermionic
determinants (denoted by $\Delta$) arising from messengers in the
fundamental representation of $SU(N_c)$. Therefore, the full one loop
effective action
$$
S_{eff}(X, \psi_X)= S_{tree}+ {\rm Tr}\, \big({\rm log}\, \Delta\big)
$$
can be written as a single trace of products of $X_{ij}$ and its
superpartner.  This means that {\it the tree level plus one loop}
contribution to
the masses of the $X$ fields must be of the form ${\rm Tr}(X^\dagger X)$, and
therefore the singlet and adjoint parts of $X$ get identical masses
through one loop.  The same is true for the fermionic partners of $X$:
at one loop the masses of the singlet $\psi_{{\rm tr}\,X}$ and
the adjoint $\psi_{X'}$ are the same.  Diagrammatically, there is a
cancellation between the tree level Weyl mass and the one loop
correction.

We note two small subtleties. First, we have assumed
here that the kinetic terms for the singlet and adjoint parts of $X$
have the same normalization.  This is true to a very good
approximation. We assumed $m \ll \Lambda \ll \Lambda_0$, which ensured
that the high-energy theory's approximate $SU(N_f)\times SU(N_f)$
symmetry is only weakly broken to $SU(N_f)_V$ at the scale $\Lambda$.
Under this larger symmetry, the singlet and adjoint transform as a
single irreducible representation, assuring equally normalized
kinetic terms, up to negligible order($\mu/\Lambda$) corrections.

Second, and irreducibly, the Goldstino is not quite $\psi_{{\rm tr}\,X}$.  As discussed in more detail in the Appendix, it
mixes slightly with the fields $\psi_{{\rm tr}\,Y}$ and $\psi_{{\rm
    tr}\,(\chi+\tilde\chi)}$, with mixing angles of order a one loop
factor, $\sim 1/16\pi^2$ and $\sim X_0/(16\pi^2\mu)$, respectively.
Consequently the tree level and one loop $\psi_X$ masses fail to
cancel precisely, though by an amount that is one further loop-order
suppressed. Thus our statement that the $\psi_X$ masses vanish at
one loop is effectively correct.

\subsection{Phenomenology of the $\gamma=0$ model}

After gauging a subgroup of the flavor group $SU(N_c)$  --- see
Eq.~(\ref{eq:symm}) --- and identifying it with the Standard Model gauge
group, the adjoint fermions $\psi_{X'}$ will carry Standard Model
gauge charges.  The fact that they are approximately massless at one
loop is unacceptable phenomenologically.  They do become massive at
two loop order, through the above-mentioned mixings, and through
explicit two loop diagrams.  For example, Standard Model gauge bosons,
which do not impact the singlet $\psi_{{\rm tr}\,X}$, generate for the
other fields a two loop mass of order
\begin{equation}
m_{\psi_{X'}} \; \sim \;g^2 \,\frac{X_0}{(16\pi^2)^2} \,\, \sim \,\, g^2\,\frac{\mu_{\phi}}{16\pi^2} \ .
\end{equation}
But the Standard Model gauginos have a one loop mass of order
${X_0 }/{16 \pi^2} \sim \mu_\phi$.  Importantly, the charge
conjugation symmetry discussed in Section \ref{sec:sqcd} forbids
significant mixing between $\lambda$ and $\psi_X$, so the masses for
the $\psi_{X'}$ fields cannot be raised through mixing effects.
{\it Consequently, requiring the gauginos are at a scale $\sim$ 1 TeV
implies the $\psi_{X'}$ would be so light that they would have already
been observed.}

\section{The deformation with $\gamma\neq 0$} \label{sec:multi}

Clearly the root of this phenomenological problem lies in treating $\psi_{X'}$ and
the Goldstino $\psi_{{\rm tr}\,X}$ on the same footing in the tree
level superpotential.  A solution is to allow non-zero $\gamma$,
\begin{equation}\label{eq:magneticdef2}
W=
h {\rm tr}(q \Phi \tilde q)-h \mu^2\,{\rm tr}\,\Phi+
\frac{1}{2}h^2 \mu_\phi \Big({\rm tr}\,(\Phi^2)+\gamma({\rm tr}\,\Phi)^2 \Big)\,.
\end{equation}
such that the two have different tree level masses. Then the total one
loop mass for $\psi_{X'}$ becomes proportional to $\gamma
\mu_\phi$.

The motivation for considering non-zero $\gamma$
\begin{equation}\label{eq:magneticdef3}
W=-h \mu^2\,{\rm tr}\,\Phi+ h {\rm tr}(q \Phi \tilde q)+\frac{1}{2}h^2 \mu_\phi \Big({\rm tr}\,(\Phi^2)+\gamma({\rm tr}\,\Phi)^2 \Big)\,,
\end{equation}
extends beyond phenomenological utility. No
symmetry enforces $\gamma=0$ once $\mu_\phi$ or
even $\mu$ are non-zero, so it is quite natural for $\gamma$ to be
nonzero.\footnote{
Considering the preserved symmetries, one might
wonder why the coefficients of $q\Phi\tilde q$ should be taken
precisely equal.
The point is that the physical couplings are constrained by
the approximate $SU(N_f)_L\times SU(N_f)_R$ in the electric theory,
which is still valid at and just below the scale $\Lambda$.  In other
words, the $\mu\to 0$ and $\mu_\phi\to 0$ limit implies equal couplings.
Nothing comparable favors $\gamma=0$.}

Let us now analyze the metastable vacua of the theory. For $h \mu_\phi
\ll \mu$ (and for $|\gamma|$ roughly of order 1), the Coleman-Weinberg
potential is approximately as in ISS. The only stable local minimum
occurs for $k=N_f-N_c$.  The multitrace deformation adds a term
proportional to the identity matrix to $W_\Phi$, so we obtain
\begin{equation}\label{eq:multi-vac1}
q_0 \tilde q_0=\mu^2-h\mu_\phi \,N_c\,\gamma\,X_0\,.
\end{equation}
\begin{equation}\label{eq:multi-vac2}
hX_0\approx \frac{\mu^2 \mu_\phi^*(1+N_c\gamma^*)}{b|\mu^2|+|\mu_\phi|^2+f(\gamma, \gamma^*)}
\end{equation}
with
$$
f(\gamma, \gamma^*)=|\mu_\phi|^2\big[N_c\, (\gamma+\gamma^*) +N_c^2\,|\gamma|^2\big]\,.
$$
In the limit $h\mu_\phi\ll \mu$, the effect of $\gamma$ is qualitatively
unimportant:
\begin{equation}\label{eq:metavac2}
h X_0  \approx\,\,\frac{\mu^2 \mu_\phi^*(1+N_c\gamma^*)}{b|\mu|^2}\;,\;q_0 \tilde q_0\approx \mu^2\,,
\end{equation}
so that $|h X_0| \gg |\mu_\phi|$.  While $\gamma\neq 0$ does not alter
the qualitative features of the vacuum, it is important, when
computing the spectrum, that the precise values (\ref{eq:multi-vac1})
and (\ref{eq:multi-vac2}) be used.

\subsection{Spectrum}\label{subsec:spectrum}

We now analyze the spectrum in the metastable vacuum.
As in Section \ref{sec:single}, the Goldstino is not massless at tree level.
Some of the one loop diagrams exactly cancel the tree level contributions
and for
this reason we discuss directly the tree level plus one loop results.

We first consider the fermions of the pseudo-modulus $X$.
The singlet fermion (the Goldstino) is massless at one loop.
For the adjoint fermions, the tree level mass $h^2 \mu_\phi$ is partially
canceled against the one loop contribution, and the full mass is of order
\beq\label{eq:psiXmassestimate}
m_{\psi_{X'}}\,\approx\,h^2 \mu_\phi \,N_c\gamma\,.
\eeq
Of course this vanishes in the limit $\gamma \to 0$, as required from
Section \ref{sec:single}.

Interestingly, we will see in Section \ref{sec:pheno} that
the Majorana gaugino masses are proportional to $(1+N_c\gamma)$.
By changing the dimensionless parameter $\gamma$, the
adjoint fermions may thus be made lighter or heavier than the gauginos.
This allows a variety of spectra with different phenomenological
signatures, see Section \ref{sec:pheno}.

As for the bosons of $X$, both the adjoint and one component of the singlet
acquire one loop masses of order $m_{CW}$; see Eq.~(\ref{eq:mcw}).
The other part of the singlet, Arg($X$), is a massive R-axion.
This is because $X$ has a large nonzero expectation value
$X_0 \sim 16 \pi^2 \muphi \gg \mu_\phi$, which spontaneously
breaks the approximate $U(1)_R$ symmetry at a scale much larger than
any explicit breaking.
The mass of the R-axion is given by
\begin{equation}\label{Raxmass}
m_a^2=\frac{2\,\sqrt{N_c}}{N_c|X_0|}\;{\rm Re}\,\big[h \mu^2\,(h^2 \mu_\phi)^* \big] \sim b |h^4\,\mu^2|\,.
\end{equation}
This is of the same order as the one loop mass $m_{CW}$, Eq.~(\ref{eq:mcw}).

Finally, the $(Y, \chi, \tilde \chi)$ and $(Z, \tilde Z, \rho, \tilde
\rho)$ sectors are as in Section \ref{subsec:review}. We remind the
reader that we have gauged the $U(1)_V$ symmetry, and $g_V$ denotes
its gauge coupling. The (otherwise massless) fields from ${\rm
tr}(\chi-\tilde \chi)$ acquire masses of order $g_V \mu$, as shown in
the table. Furthermore, the NG bosons
from $(\rho, \,\tilde \rho,\,Z,\,\tilde Z)$
acquire a one loop mass of order $g_{SM} \mu/4\pi$ once the Standard Model
is gauged, as a subgroup of the flavor
symmetry group. The lightest of these
is stable due to the unbroken messenger number $U(1)'$ from
Eq.~(\ref{eq:breaking}).

\begin{figure}[!t]\begin{center}
\includegraphics[scale=0.80]{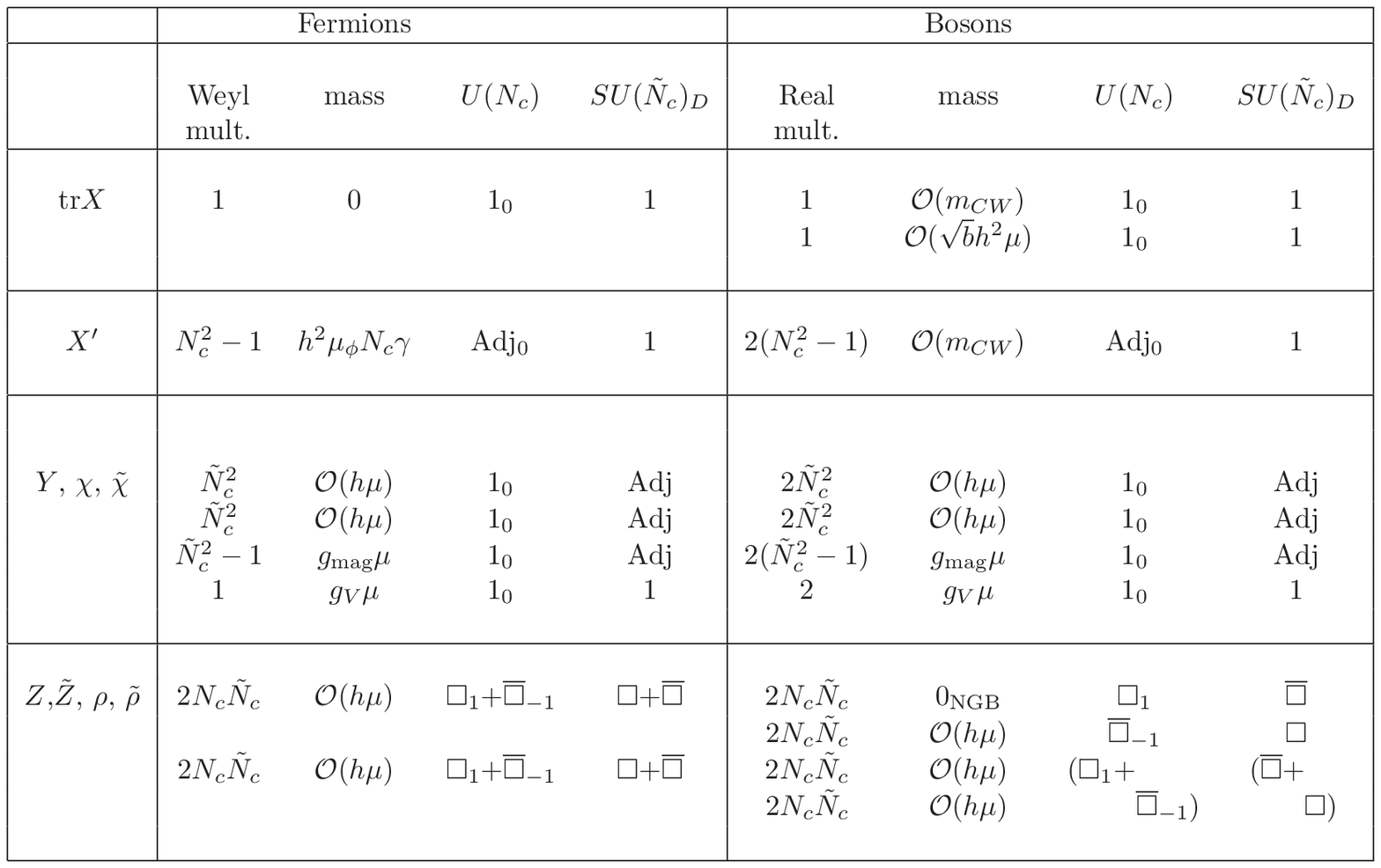}
\caption{The mass spectrum, including one loop corrections (but
  without Standard Model gauge interactions), grouped in sectors with
  ${\rm Str} \, M^2=0$.  Notice the appearance of the Goldstino in the
  ${\rm tr}\,(X)$ sector.  The details of the spectrum are described
  further in the text.  Notation is as in Figure 1.}
\label{table:oneloopmasses}
\end{center}
\end{figure}

\subsection{Lifetime of the metastable vacuum}\label{subsec:vacuumlifetime}

Here we check that the metastable non-supersymmetric vacuum can be
sufficiently long-lived. This vacuum can decay to the ISS-like
supersymmetric vacuum with $k=N_f-N_c$, or to the supersymmetric vacua
with $k<N_f-N_c$ (see Section \ref{subsec:magnetic}). The decay to
the vacua with $k<N_f-N_c$ requires changing the expectation value
of (some of the elements of) $q \tilde q$, from $h\mu^2$ to $0$. This
is strongly suppressed by the quartic potential term $V=\ldots+ |h q
\tilde q|^2$. The dominant decay channel will be to the supersymmetric
vacuum with $k=N_f-N_c$, which we now analyze.

The lifetime of the vacuum may be estimated using
semiclassical techniques and is proportional to the exponential
of the bounce action, $e^B$ \cite{Coleman:1977py}. We will see that the tunneling takes place in the direction 
of ${\rm tr}\,X$, in a region where $q \tilde q \approx \mu^2$ is almost constant. 
The potential as a function of ${\rm tr}\,X$, including the one loop quantum
corrections from the Coleman-Weinberg potential, is given in the Appendix and
shown in Figure \ref{fig:Vcw-slice}.
It may be modeled as a triangular barrier, and the bounce action may be
estimated using the results in \cite{Duncan:1992ai}.

We will see in the next section that, in order
to have large enough gaugino masses but a low SUSY-breaking scale and
low sfermion masses, the ratio $\muphi/\mu$ cannot be made too small.
Nonetheless, it is useful to first analyze the bounce action in the
limit $\muphi\ll \mu$, where it is clear the vacuum is parametrically
stable.

The dimensionful parameters controlling the shape of the potential are $\mu$ and
$\muphi$. We assume $h$, $\gamma$, $N_f$, and $N_c$ are all of order 1.  
The SUSY vacua are parametrically far away from the metastable vacua in the limit
\beq\label{eq:epsilon}
\epsilon \equiv \Big|\frac{\muphi^2}{b\mu^2}\Big| \ll 1 \ .
\eeq
In this limit, the calculation of the bounce action is very similar to that done in
\cite{est}, as long as only ${\rm tr}\, X$ varies.  Let us assume $q\tilde q$
is essentially constant.

The metastable SUSY-breaking vacuum lies at $X_0\sim\muphi/b$,
the peak of the potential is near $X_{\textrm{\footnotesize{peak}}}\sim b\mu^2/\muphi$, and
the SUSY vacuum is at $X_{\textrm{\footnotesize{susy}}}\sim\mu^2/\muphi$, where
phases and $\mathcal{O}(1)$ numbers have been ignored.
Moreover, the potential difference between the peak and the metastable
SUSY-breaking minimum is roughly
$V(X_{\textrm{\footnotesize{peak}}})-V(X_0)\sim b\,\mu^4$, much smaller than $V(X_0)-V(X_{\textrm{\footnotesize{susy}}}) \sim \mu^4$.
The results of \cite{Duncan:1992ai} 
then show that 
the field tunnels not to
the SUSY vacuum directly but rather to $X_{\textrm{\footnotesize{tunnel}}}
\gtrsim X_{\textrm{\footnotesize{peak}}}$.
For this value of $X_{\textrm{\footnotesize{tunnel}}}$,  Eq.~(\ref{eq:F-term1}) implies $q \tilde q \approx \mu^2$, 
and thus $q \tilde q$ indeed stays approximately constant in the tunneling region. 
This confirms that the results in \cite{Duncan:1992ai} apply.

In the limit $\epsilon \ll 1$, the bounce action scales parametrically as
\beq\label{eq:bounceaction}
B \sim
    \frac{(X_{\textrm{\footnotesize{tunnel}}})^4}{V(X_{\textrm{\footnotesize{peak}}})-V(X_0)}
  \sim b \, \frac{1}{\epsilon^2},
\eeq
where we have neglected some numerical factors, see \cite{Duncan:1992ai}.  
Thus, $B\rightarrow\infty$ as $\epsilon\rightarrow 0$, and the
metastable vacuum can be made parametrically long-lived.

In Section \ref{sec:pheno}, we will see that in order to obtain sfermion masses
that are roughly of the same size as gaugino masses, we need to take $\muphi \sim b\mu$ 
(and thus $\epsilon\sim b$.)  In this regime $X_0$,
${X_{\textrm{\footnotesize{peak}}}}$
and ${X_{\textrm{\footnotesize{tunnel}}}}$ are all parametrically
of order 
$b{X_{\textrm{\footnotesize{SUSY}}}}$.
A numerical study is required to determine the existence and lifetime
of the metastable vacuum.  Taking the gaugino masses to lie at their experimental
lower bound, of order 100 GeV, we find that the existence of a metastable vacuum
sets a lower bound on the sfermion masses --- typically a few TeV for the squarks and
at least a few hundred GeV for the right-handed sleptons.  Once such a metastable
vacuum is obtained, it is easy to make the bounce action larger than 
the required 400 by a small 
increase (of order 5\%) in the sfermion masses. The details of the spectrum, together
with a more precise estimate of the lower bound on the sfermion masses, and the
implications for the tuning of electroweak symmetry breaking, will be given in
\cite{EFSST2}.

\section{Comments on the phenomenology} \label{sec:pheno}

This section briefly discusses some of the phenomenology
associated with the multitrace deformation of the ISS model,
equation (\ref{eq:magneticdef2}).
The details will be left to a forthcoming publication \cite{EFSST2}.

The ISS-like supersymmetry breaking models are interesting from a
phenomenological point of view due to the presence of the large global
symmetry group
\beq\label{eq:unbrokenglobalsym}
SU(\tilde N_c)_V \times SU(N_c) \times U(1)'\,.
\eeq
A model of direct gauge mediation can be built by weakly gauging a subgroup of
(\ref{eq:unbrokenglobalsym}) and identifying it with the Standard Model (SM)
gauge group.
The fields $\rho$, $Z$, $\tilde{\rho}$, and $\tilde{Z}$ in (\ref{eq:paramPhi}) and
(\ref{eq:paramq}) act as messengers that mediate the supersymmetry breaking
effects to the visible sector.
Loops involving these messengers can give non-zero masses to the scalar superpartners
of the SM fermions and, provided there is no unbroken R-symmetry, non-zero
Majorana fermion masses to the gauginos.

In this section, we will consider gauging the $SU(3)\times SU(2) \times U(1)$ subgroup of
$SU(N_c)$ for $N_c = 5$ in the $\gamma\neq 0$ model, and identifying it
with the SM gauge group. (The effect of gauging a subgroup of $SU(\tilde{N}_c)_V$ will be discussed in \cite{EFSST2}.)

Under the SM gauge group $SU(3)_C\times SU(2)_L\times U(1)_Y$, the adjoint field $X'$
decomposes as
\beq\label{eq:Xdecomposition}
X' = X_{\bf 24} = X_{({\bf 8},{\bf 1})_0}
           \oplus X_{({\bf 1},{\bf 3})_0}
           \oplus X_{({\bf 3},{\bf 2})_{-5/6}}
           \oplus X_{({\bf \bar{3}},{\bf 2})_{5/6}}
           \oplus X_{({\bf 1},{\bf 1})_0}.
\eeq
The fermions from the superfields ${X_{({\bf 8},{\bf 1})_0}}$,
${X_{({\bf 1},{\bf 3})_0}}$, and ${X_{({\bf 1},{\bf 1})_0}}$ carry the
same gauge charges as the gluino, wino, and bino, respectively, and
the first two could be directly
produced at colliders.\footnote{The $X$ bosons in
(\ref{eq:Xdecomposition}) get a mass of order $\sqrt{b}h^2\mu \sim
\mc{O}$(10 TeV) from the Coleman-Weinberg potential and are thus
rather heavy. If produced in the early Universe, they would have
decayed promptly into $\psi_X$ and a gaugino, excepting gauge singlets
which would decay a bit more slowly through higher dimension
operators.} \ Also, there are new
light fermions from the superfields $X_{({\bf 3},{\bf 2})_{-5/6}}$ and
$X_{({\bf \bar{3}},{\bf 2})_{5/6}}$; these are stable unless given new
interactions, and require a special discussion below.

\subsection{Phenomenology of $\psi_{X'}$ and $\lambda$}\label{sec:psiX_gaugino_pheno}

A very important property of the model is that {\it the gauginos and the adjoint $\psi_{X'}$
do not mix}.  This is due to the fact that $\lambda$ and $\psi_{X'}$ have
charge conjugation transformations that differ by a sign,
\beq
C(\psi_{X'_{ij}})=\psi_{X'_{ji}}\;,\;C(\lambda_{ij})=-\lambda_{ji}\,.
\eeq
This discrete symmetry forbids any mixing at low orders
between the two sets of fermions.  More precisely, $C$-violation in
the SM allows $\lambda$ and $\psi_{X'}$ to mix, but this occurs only at
three loops and is thus negligibly small.

\vskip 2mm

Let us estimate the gaugino and $\psi_{X'}$ masses.
As discussed in Section \ref{subsec:spectrum}, the metastable vacuum has an approximate
R-symmetry that is spontaneously broken through the non-zero vev $X_0 \sim(1+N_c\gamma)\muphi/b$,
where $b\sim 1/(16\pi^2)$ is a loop factor (\ref{eq:b}).
Therefore, gauginos obtain a one loop mass of order
\beq\label{eq:gauginomass}
m_\lambda \sim \frac{g^2}{16\pi^2}\,X_0 \sim g^2\, (1+N_c\gamma)\muphi\,.
\eeq
Neglecting $\mc{O}(1)$ numbers and factors of the gauge coupling $g$, an interesting phenomenology is obtained for
\beq
m_\lambda \sim \mc{O}\,{\textrm{(1 TeV)}}\,,
\eeq
i.e.~for
\beq\label{eq:muphiestimate}
\muphi\sim\mc{O}(1 \;\rm{TeV}).
\eeq
The $\psi_{X'}$ also obtain a mass at one loop, which, using equation (\ref{eq:psiXmassestimate}), is of order
\beq\label{eq:psiXmass}
m_{\psi_{X'}} \sim h^2 \, \muphi \, N_c \, \gamma \sim  \gamma \, \times \, \mc{O}\,{\textrm{(1 TeV)}}\,,
\eeq
neglecting factors of $h$ and $g$ and other $\mc{O}(1)$ numbers.
By adjusting $\gamma$, $\psi_{X'}$ can be made heavier or lighter
than $\lambda$, leading to very different collider signatures as we will discuss
next.

\vskip 2mm

The $\psi_{X'}$ do not mix with the Standard Model gauginos at a level
that determines their decays.  Instead, if they are heavy enough, they
can decay (promptly) into a gaugino and a gauge boson through the
dimension five operator $\psi_{X'} \sigma^{\mu\nu} \lambda F_{\mu\nu}$:
\beq \psi_{X'} \to \lambda + {\textrm{gauge boson}}\,.  \eeq
The gauginos can decay through all the usual supersymmetric
decay modes, and/or through the standard coupling of each gaugino to a gauge boson and
Goldstino:
\beq \lambda \to \psi_{{\rm tr}\,X} +
{\textrm{gauge boson}}\,  \eeq

If instead the $\psi_{X'}$ are lighter than the gauginos, then
the gauginos will decay into the $\psi_{X'}$ plus a gauge boson via
the above-mentioned operator.
The $\psi_{X'}$ decays to a gauge boson and an off-shell gaugino.
The precise decay modes and the lifetime of the $\psi_{X'}$ depend on the
details of the spectrum, and will be discussed further in \cite{EFSST2}.

\vskip 2mm

From (\ref{eq:Xdecomposition}), we see that there are
\emph{new $({\bf 3,2})$ fermions}, with charges
$({\bf 3},{\bf 2})_{-5/6}$ and $({\bf \bar{3}},{\bf 2})_{5/6}$.
By binding to quarks, these form hadrons, some of which are charged.
The lightest of these novel hadrons, whether
charged or neutral, would be stable in the model as described so far.  But
this would be ruled out, since these hadrons would have
been created in the early Universe, violating the bounds on the
existence of heavy stable
particles~\cite{Smith:1982qu,Perl:2001xi}. These fermions must thus be
made to decay through additional baryon-number violating
operators in the superpotential and/or the K\"ahler potential.
In~\cite{EFSST2}, we will show that additional dimension five K\"ahler
potential terms, coupling the adjoint $X'$ to SM quarks and leptons,
can allow the $({\bf 3},{\bf 2})$ fermions to decay without affecting
Big-Bang Nucleosynthesis or violating current bounds on proton decay.

\subsection{Sfermion masses, the SUSY-breaking scale and a light gravitino}

Since the supersymmetry breaking scale is $|\sqrt{F}|=|\sqrt{h} \mu|$ and the mass scale of the messengers is of the
same order, the soft scalar masses are roughly given by
\beq\label{eq:sfermionmass}
m_S \sim \frac{g^2}{16\pi^2}\,\mu\,.
\eeq
Comparing this to (\ref{eq:gauginomass}), the sfermions and gauginos have similar masses if
\beq\label{eq:muphivalue}
\mu_\phi \sim \mu/(16 \pi^2).
\eeq
We recall that the existence and longevity of the metastable vacuum requires $\muphi \ll\mu$, see Section \ref{subsec:vacuumlifetime}.

More concretely, there is an interesting parameter region characterized by (\ref{eq:muphivalue}) and a low supersymmetry breaking scale
\begin{equation}\label{eq:parameter}
\sqrt{F} \approx \mu \sim \mathcal O \,( 100 - 200\; {\rm TeV})\,.
\end{equation}
In this case, one can show (see~\cite{EFSST2}) that the heaviest sfermions (squarks) have masses of a few TeV, the lightest sfermions (right-handed sleptons) haves masses of a few
hundred GeV, the gaugino masses are of order several hundred GeV, and there is a large enough lifetime for the metastable vacuum. The gravitino mass is
\beq
m_{3/2} \sim \frac{F}{\sqrt{3}M_{\rm Pl}} \sim \mc{O}({\textrm{1--10 eV}})\,,
\eeq
where $M_{\rm Pl}\simeq 2.4\times 10^{18}$ GeV is the reduced Planck mass.
Such a light gravitino does not violate any cosmological or astrophysical
constraints \cite{Viel:2005qj}.

\subsection{Further comments on the spectrum}

As discussed in Section \ref{sec:multi}, the messenger sector $(\rho,
\,\tilde{\rho}, \,Z, \,\tilde{Z})$ contains $2N_c\tilde N_c$
real NG bosons, all of which become massive at one loop after weakly gauging the flavor
symmetry. In the parameter range (\ref{eq:parameter}), this mass is of
order of several TeV. The $U(1)'$ messenger number in
(\ref{eq:breaking}) forbids the decay of the lightest of these
messenger particles, which is thus stable.
If the lightest messenger is
neutral and weakly interacting and has an appreciable relic density, it
would have a tree-level coupling to nuclei via $Z$-boson exchange and would have been
seen at a dark matter direct detection experiment
\cite{Dimopoulos:1996gy}-\cite{Angle:2007uj}.
If the stable state is charged and/or colored, the experimental constraints
are even stronger \cite{Smith:1982qu,Perl:2001xi}.
Thus experimental constraints rule out the possibility that the lightest messenger
is dark matter; this will be investigated further in~\cite{EFSST2}.

We also note that the SM gauge couplings have a Landau pole well below
the GUT scale, due to the presence of extra matter charged under the
SM gauge group.  As one runs up to the high scale, the $SU(3)_C$ gauge
coupling blows up first at about $10^9$ $(10^7)$ GeV for
$\tilde{N_c}=1$ $(3)$, so that new physics has to enter at or below this
scale. Larger values of $\tilde{N_c}$ lower this scale to the point
that it affects our discussion materially.
See~\cite{Abel:2008tx} for a recent discussion of the Landau pole
problem in ISS-like SUSY-breaking models.

\subsection{Illustrative choices of parameters}

We postpone a careful study of the various constraints to
\cite{EFSST2}, but preliminarily it appears possible to satisfy
simultaneously all of the conditions considered above.  For example,
for $\tilde{N_c}=1$,\footnote{In this case, the magnetic gauge group is trivial and, after a field redefinition, the superpotential is given by (\ref{eq:magneticdef}) plus ${\rm det}\,\Phi/\Lambda^{N_c-2}$. For $N_c>2$ this term is negligible near the origin, so our analysis is self-consistent.} the parameters of the electric theory
Eq.~(\ref{eq:electricdef}) that are consistent with (\ref{eq:muphivalue}) and
(\ref{eq:parameter}) are $m$ of order 0.01--10 TeV,
$\Lambda\sim 10^{3-5}$ TeV, and $\Lambda_0\sim 10^{6-9}$ TeV. With these choices, the models appear to
have no insuperable problem below the scale of the Landau pole.

On the other hand, for $\tilde N_c \ge 3$, $\Lambda$ has to be below
$10^3$ TeV, and the ratio $m/\Lambda$ is not parametrically small. In
this case, the corrections from the microscopic theory are not
guaranteed to be small, and the violations of the approximate
symmetries may be large.  In particular, the cancellations described
in section \ref{subsec:lightf} may be imperfect, requiring a more
elaborate analysis.  However, the argument for nonzero $\gamma$ still
holds, and its effects can still dominate, in which case the
phenomenology outlined here will be largely unchanged.

\subsection{Summary}

While these models are not yet entirely plausible, they represent an
advance over the models with $SU(N_c)$ gauged and $\gamma=0$, which as
we showed are excluded by the presence of overly-light charged and
colored fermions.  We have demonstrated that with $\gamma\neq 0$, it
is possible to obtain models with a long-lived metastable vacuum, a
spectrum with all standard model superpartners in the TeV range, and
with no obvious unresolvable conflict with any experiment.

The minimal versions of these models have new TeV-scale fermions in the adjoint
representations of the Standard Model gauge group that do not mix with
standard model gauginos.  They also have squarks and sleptons
significantly heavier than the gauginos, and exotic stable hadrons
which must be made to decay through additional interactions.  They
also suffer from the ubiquitous intermediate-scale Landau pole for
standard model gauge couplings.  We will pursue various associated
model-building issues, and study in more detail the phenomenology of
these models
in \cite{EFSST2}.

\subsection*{Acknowledgments}
We would like to thank Evgeny Andriyash, Masahiro Ibe, Paul Langacker, David Shih,
and especially Scott Thomas for valuable discussions.
This research was supported by DOE grant DE-FG01-06ED06-02.  R.E. is supported by the
US DOE under contract number DE-AC02-76SF00515.
K.S. is supported by NSF grant PHY-0505757 and the University of Texas A\&M.
G.T. would like to thank KITP for their hospitality and support through the NSF Grant PHY05-51164.


\appendix


\section{One loop calculations}\label{app:mass}

In this appendix we collect the one loop calculations for the ISS model with multitrace quadratic deformations. The superpotential is
\begin{equation}
W=h\,\mbox{tr}\,q\Phi\tilde{q}-h\mu^2\,\mbox{tr}\,\Phi+\frac{1}{2}h^2\mu_\phi\,\mbox{tr}\,\Phi^2+\frac{1}{2}h^2\mu_\phi\gamma\left(\mbox{tr}\,\Phi\right)^2
\end{equation}
where $\Phi=\Phi_{N_f\times N_f}$, $q=q_{\tilde{N}_c\times N_f}$ and $\tilde{q}=\tilde{q}_{N_f\times\tilde{N}_c}$.

\subsection{Messenger sector}\label{app:mess}

Let us consider separately the cases $k=N_f-N_c$ and $k< N_f-N_c$ (see Section \ref{subsec:k-cases}).

\subsubsection*{Case $k=N_f-N_c$}
The parametrization of the metastable minima is given by Eqs.~(\ref{eq:paramPhi}) and (\ref{eq:paramq}).
Around these minima the superpotential is
\begin{eqnarray}\label{eq:appW}
W &=& hq_0 \tilde q_0\,\mbox{tr}\,Y-h\mu^2\,\mbox{tr}\,Y-h\mu^2\,\mbox{tr}\,X+h\,\mbox{tr}\,q_0Y\tilde{\chi}+h\tilde{q}_0\,\mbox{tr}\,\chi Y\nonumber\\
 && +hq_0\,\mbox{tr}\,Z\tilde{\rho}+h\tilde{q}_0\,\mbox{tr}\,\rho\tilde{Z}+\frac{1}{2}h^2\mu_\phi\left(\mbox{tr}\,Y^2+\gamma(\mbox{tr}\,Y)^2\right)+h^2\mu_\phi\,\mbox{tr}\,Z\tilde{Z}\nonumber\\
 && +\frac{1}{2}h^2\mu_\phi\left(\mbox{tr}\,X^2+\gamma(\mbox{tr}\,X)^2\right)+h^2\mu_\phi\gamma\,\mbox{tr}\,X\,\mbox{tr}\,Y\nonumber\\
 && +h\,\mbox{tr}\,\chi Y\tilde{\chi}+h\,\mbox{tr}\,\rho X\tilde{\rho}+h\,\mbox{tr}\,\rho\tilde{Z}\tilde{\chi}+h\,\mbox{tr}\,\chi Z\tilde{\rho}
\end{eqnarray}
and the non-zero F-term is
\begin{equation}
\partial_{X_{ij}}W=\left(-h\mu^2+h^2\mu_\phi(1+N_c\gamma)X_0\right)\delta_{ij}\,.
\end{equation}
We recall the ansatz (\ref{eq:X0}),
\begin{equation}\label{eq:X02}
\langle X\rangle= X_0 \,I_{N_c \times N_c}\;,\;
\langle \chi \rangle=q_0\,I_{\tilde N_c \times \tilde N_c}\;,\;
\langle \tilde \chi \rangle=\tilde q_0\,I_{\tilde N_c \times \tilde N_c}\,.
\end{equation}
The $q_0\tilde{q}_0$ vev completely Higgses the dual gauge group $SU(\tilde{N}_c)_G$ and the $U(1)_V$.  To determine $X_0$, one must compute the Coleman-Weinberg potential from the tree level masses of the messenger sector.  The ansatz (\ref{eq:X02}), which will be checked self-consistently, simplifies the computations since the mass eigenstates are then independent of their flavor index.  One can thus suppress color and flavor indices in the following.

The messenger sector contains the fields $\rho$, $\tilde{\rho}$, $Z$ and $\tilde{Z}$, that couple to the non-zero F-term.  Let us define
\begin{equation}
\hat{\psi}=\left(\;\psi_{\rho}\;\;\;\psi_{Z}\;\right)^T\;\;\;\;\;\;\hat{\tilde{\psi}}=\left(\;\psi_{\tilde{\rho}}\;\;\;\psi_{\tilde{Z}}\;\right)^T\;\;\;\;\;\;\hat{\phi}=\left(\;\rho\;\;\;Z\;\;\;\tilde{\rho}^*\;\;\;\tilde{Z}^*\;\right)^T
\end{equation}
for the messenger gauge eigenstates.  The Weyl fermions combine into Dirac fermions and the messenger masses can be written as
\begin{equation}
\mathcal{L}_{mess,mass}=-\hat{\tilde{\psi}}M_{mess,f}\hat{\psi}-h.c.-\hat{\phi}^\dagger M_{mess,b}^2\hat{\phi}
\end{equation}
where the messenger mass matrices are
\begin{equation}
M_{mess,f}=h\left(
\begin{array}{cc}
X_0 & q_0\\
\tilde{q}_0 & h\mu_\phi
\end{array}
\right),\;\;M_{mess,b}^2=\left(
\begin{array}{cc}
M_{mess,f}^\dagger M_{mess,f} & -h^*F_X^*\\
-hF_X & M_{mess,f}M_{mess,f}^\dagger
\end{array}
\right)
\end{equation}
and
\begin{equation}
-F_X^*=h\left(
\begin{array}{cc}
-\mu^2+h\mu_\phi(1+N_c\gamma)X_0 & 0\\
0 & 0
\end{array}
\right).
\end{equation}
For $\tilde{q}_0=q_0$ the fermionic and bosonic messenger masses are ($\sigma=\pm1$ and $\eta=\pm1$)
\begin{eqnarray}\label{nonSUSYmasses}
m^2(X_0)&&= |h|^2\left(|q_0|^2+\frac{1}{2}|X_0|^2+\frac{1}{2}|h\mu_\phi|^2\right.\\ &&\left.+\frac{1}{2}\sigma\sqrt{\left(|X_0|^2-|h\mu_\phi|^2\right)^2+4|q_0X_0^*+q_0^*h\mu_\phi|^2}\right)\nonumber\\
\tilde{m}^2(X_0)&& = |h|^2\left(|q_0|^2+\frac{1}{2}|X_0|^2+\frac{1}{2}|h\mu_\phi|^2+\frac{1}{2}\eta|\mu^2-h\mu_\phi(1+N_c\gamma) X_0|\right.\\
 && \left.+\frac{1}{2}\sigma\sqrt{\left(|X_0|^2-|h\mu_\phi|^2+\eta|\mu^2-h\mu_\phi(1+N_c\gamma) X_0|\right)^2+4|q_0X_0^*+q_0^*h\mu_\phi|^2}\right)\nonumber.
\end{eqnarray}
The fermion masses have multiplicity $4N_c\tilde{N}_c$ while the complex boson masses have multiplicity $2N_c\tilde{N}_c$.

The messenger mass matrices can be diagonalized by unitary matrices $U_f$, $\tilde{U}_f$ and $U_b$ such that
\begin{equation}
\psi=U_f\hat{\psi}\;\;\;\;\;\;\tilde{\psi}=\tilde{U}_f\hat{\tilde{\psi}}\;\;\;\;\;\;\phi=U_b\hat{\phi}
\end{equation}
where $\psi$, $\tilde{\psi}$ and $\phi$ are messenger mass eigenstates.  The quadratic lagrangian for the messengers is therefore of the canonical form
\begin{multline}
\mathcal{L}_{mess}=-\sum_{a=1}^4\phi_a^\dagger\left(D^2+\tilde{m}_a^2\right)\phi_a\\
+\sum_{a=1}^2\left(\bar{\psi}_ai\bar{\sigma}^\mu D_\mu\psi_a+\bar{\tilde{\psi}}_ai\bar{\sigma}^\mu D_\mu\tilde{\psi}_a-m_a(\tilde{\psi}_a\psi_a+\bar{\tilde{\psi}}_a\bar{\psi}_a)\right).
\end{multline}
Due to the charge conjugation symmetry, it is possible to write the mixing matrices such that $(U_b)_{a\{1,2\}}=(U_b)_{a\{3,4\}}^*$ and $\tilde{U}_f=U_f$.  This can be easily seen from the mass matrices for $\tilde{q}_0=q_0$.  This property will be useful when computing one loop corrections to light masses.

\subsubsection*{Case $k<N_f-N_c$}
The fluctuations are parametrized as in Eq.~(\ref{eq:param}), so there are extra messenger superfields $(\varphi, \tilde \varphi)$. The analysis of $(\rho,\, \tilde{\rho}, \,Z,\,\tilde{Z})$ proceeds along the same lines as in the case $k=N_f-N_c$, except that the fermion messenger masses have now multiplicity $4(N_f-k)k$ while the complex boson messenger masses have multiplicity $2(N_f-k)k$.

The masses of $\varphi$ and $\tilde{\varphi}$ are ($\eta=\pm1$)
\begin{eqnarray}\label{eq:mvarphi}
m_\varphi^2(X_0) &=& |hX_0|^2\nonumber\\
\tilde{m}_\varphi^2(X_0) &=& |h|^2\left(|X_0|^2+\eta|\mu^2-h\mu_\phi X_0|\right).
\end{eqnarray}
The fermion masses have multiplicity $4(N_f-k)(\tilde{N}_c-k)$ while the complex boson masses have multiplicity $2(N_f-k)(\tilde{N}_c-k)$.  Importantly, in the limit of small deformation, (\ref{eq:mvarphi}) forces $|X_0|\gtrsim|\mu|$ to avoid tachyons.

\subsection{One loop bosonic action}

The tree level pseudo-moduli are given by $X_0$ and ${\rm Re}\,{\rm tr}(\chi-\tilde \chi)$, and they are stabilized by one loop contributions. For $\mu_\phi \ll \mu$, the one loop effective potential for ${\rm Re}\,{\rm tr}(\chi-\tilde \chi)$ is the same as in~\cite{iss} (see Eq.~(\ref{eq:Vcw2}).) As a result, this field is stabilized at the origin and acquires a mass of order $|h^4 \mu^2|/(8\pi^2)$.

Let us now analyze the pseudo-modulus $X_0$; for $k \le N_f-N_c$, this is a $(N_f-k)\times(N_f-k)$ matrix. The ISS-type vacua correspond to $k=N_f-N_c$. We will argue here that the new metastable vacua corresponding to the case $k<N_f-N_c$ do not exist, as they are located in a region where some of the fields become tachyonic.  The only remaining metastable vacua will be the ISS-type vacua.

The one loop correction from integrating out the messenger fields is
\begin{eqnarray}
V_{CW} &&= \frac{(N_f-k)k}{32\pi^2}\sum_{\sigma,\eta=\pm1}\left[\tilde{m}(X_0)^4\log\frac{\tilde{m}(X_0)^2}{\Lambda^2}-m(X_0)^4\log\frac{m(X_0)^2}{\Lambda^2}\right]\\
 +&& \frac{(N_f-k)(\tilde{N}_c-k)}{32\pi^2}\sum_{\eta=\pm1}\left[\tilde{m}_\varphi(X_0)^4\log\frac{\tilde{m}_\varphi(X_0)^2}{\Lambda^2}-m_\varphi(X_0)^4\log\frac{m_\varphi(X_0)^2}{\Lambda^2}\right].\nonumber
\end{eqnarray}
with masses given in Section \ref{app:mess}. We find that the full potential
\begin{equation}
V=V_{tree}+V_{CW}
\end{equation}
has a metastable vacuum if $k=N_f-N_c$, but there are no metastable vacua for $k <N_f-N_c$. Let us discuss in more detail how this occurs.

For $k=N_f-N_c$, the messengers are non-tachyonic
for any $X_0$; see Eq.~(\ref{nonSUSYmasses}).
As explained in Section \ref{subsec:k-cases}, the
metastable vacuum appears because quantum corrections at small $X_0$
are large enough
to overwhelm the slope of the classical potential, which
would otherwise push $X_0$ toward the supersymmetric
vacua. The supersymmetry breaking vacuum is located in the range $|X_0
/ \mu| \lesssim 1$, far from the supersymmetric vacuum.

The situation for $k<N_f-N_c$ is very different, because the
messengers $(\varphi, \tilde \varphi)$ are tachyonic at small $X_0$;
see Eq.~(\ref{eq:mvarphi}).  For $|X_0 / \mu| \gtrsim 1$ these
tachyons are absent, but in this
regime the one loop corrections $V_{CW}(X_0)$ grow only
logarithmically with $|X_0|$, and cannot compete with the classical
potential to create a metastable vacuum. One may directly check that
the Hessian of the potential always has a negative eigenvalue for
$|X_0|\gtrsim|\mu|$ (and all values of $k$). Notice that if one used
the quadratic expansion of $V_{CW}$ around the origin $X_0/\mu=0$,
instead of the full logarithmic form, it would suggest the existence
of metastable vacua with $k<N_f-N_c$ and $|X_0/\mu| \sim
1$~\cite{kutasov}.  But this approximation is inconsistent, and when
the full logarithmic dependence of $V_{CW}$ is included, these vacua
become unstable and disappear.

\vskip 2mm

Summarizing, only the ISS-type minima with $k=N_f-N_c$ survive, and the adjoint ($X'$) and singlet (${\rm tr}\,X$) components of the pseudo-modulus $X$ acquire one loop masses
\begin{eqnarray}
m_{X'}^2 &\approx& b\,|h^2\mu|^2+|h^2\mu_\phi|^2\\
m_{{\rm tr}\,X}^2 &\approx& b\,|h^2\mu|^2+|h^2\mu_\phi(1+N_c\gamma)|^2.
\end{eqnarray}
The R-axion, discussed in section \ref{subsec:spectrum}, has a mass of
this same order, Eq.~(\ref{Raxmass}).  All bosons which were light at
tree level thus become heavy at one loop, with masses of order
$m_{CW}=\sqrt{b}\,|h^2\mu|$.

\subsection{One loop fermionic action}

In this section we discuss the low energy fermionic spectrum of the theory, taking into account one loop effects.

\subsubsection*{Goldstino}

At one loop, the Goldstino appears as a combination of $\psi_{{\rm tr}\,X}$, $\psi_{{\rm tr}\,Y}$ and $\psi_{{\rm tr}\,(\chi+\tilde \chi)}$, which we now determine. The charge conjugation symmetry forbids mixings with $\psi_{{\rm tr}\,(\chi-\tilde \chi)}$, which is eaten by the $U(1)_V$ gauge fermion and has mass $g_V \mu$.

First, at tree level, in the limit $\mu_\phi=0$, $\psi_{{\rm tr}\,Y}$ and $\psi_{{\rm tr}\,(\chi+\tilde \chi)}$ form a Dirac fermion of mass $h\mu$, while $\psi_{{\rm tr}\,X}$ is massless; see Eq.~(\ref{eq:appW}). When $\mu_\phi$ and $\gamma$ are nonzero, $\psi_{{\rm tr}\,X}$ acquires a mass term proportional to $\mu_\phi$, and there is a $\psi_{{\rm tr}\,X}$-$\psi_{{\rm tr}\,Y}$ mixing of order $\gamma \mu_\phi$. There is no linear combination of the fields $\psi_{{\rm tr}\,Y}$, $\psi_{{\rm tr}\,(\chi+\tilde \chi)}$ and $\psi_{{\rm tr}\,X}$ that is massless at tree level.

Once one loop effects are taken into account, supersymmetry is spontaneously broken, so we should get a massless Goldstino. Since the dominant F-term comes from $F_{{\rm tr}\,X}$, the Goldstino will be approximately aligned with $\psi_{{\rm tr}\,X}$. Indeed, the tree level plus one loop $\psi_{{\rm tr}\,X}\,\psi_{{\rm tr}\,X}$ mass element is (using the messenger mass eigenbasis),
\begin{equation}\label{eq:loop-mG}
m_{\psi_{{\rm tr}\,X}} = h^2\mu_\phi(1+N_c\gamma)-\frac{2h^2\tilde{N}_c}{16\pi^2}\,\sum_{j=1}^4\sum_{k=1}^2\, (U_f^*)_{k1}\,(\tilde U_f^*)_{k1}\,(U_b^*)_{j1}\,(U_b)_{j3}\,I[\tilde m_j,m_k]
\end{equation}
where the sums are over messenger fields and
\begin{equation}\label{eq:I-def}
I(\tilde m_j,m_k)=m_k \left[\ln\left(\frac{\Lambda^2}{m_k^2}\right)-\frac{\tilde{m}_j^2}{\tilde{m}_j^2-m_k^2}\ln\left(\frac{\tilde{m}_j^2}{m_k^2}\right)\right].
\end{equation}
It can be checked that the tree and one loop terms in (\ref{eq:loop-mG}) largely cancel, leaving only a term of order $\mu_\phi/(16\pi^2)$, of the same size as two loop corrections.

There are also one loop mixings between $\psi_{{\rm tr}\,X}$ and $\psi_{{\rm tr}\,Y}$, $\psi_{{\rm tr}\,(\chi+\tilde \chi)}$. For simplicity, let us consider first the ISS model, corresponding to the limit $\mu_\phi=0$. The mass-mixing comes from the two-point function $\psi_{{\rm tr}\,X}\,\psi_{{\rm tr}\,(\chi+\tilde \chi)}$, which is allowed by R-symmetry. A calculation along the same lines as in (\ref{eq:loop-mG}) shows that this mass-mixing is of order $\mu/(16\pi^2)$. The Goldstino is hence predominantly in the $\psi_{{\rm tr}\,X}$ direction, with a small (of order $1/(16\pi^2)$) component along $\psi_{{\rm tr}\,Y}$. This implies that in ISS, one loop corrections generate a nonzero F-term
$$
|F_{{\rm tr}\,Y}| \sim \frac{|F_{{\rm tr}\,X}|}{16\pi^2}\,.
$$

For $\mu_\phi/\mu$ nonzero but small, the Goldstino also has a small component along $\psi_{{\rm tr}\,(\chi+\tilde\chi)}$, with mixing angle of order $|X_0/(16\pi^2\mu)|$. This is smaller than the mixing of $\psi_{{\rm tr}\,X}$ and $\psi_{{\rm tr}\,Y}$, and is consistent with a one loop F-term
$$
|F_{{\rm tr}\,(\chi+\tilde \chi)}| \sim \Big|\frac{X_0}{16\pi^2 \mu}\Big|\,|F_{{\rm tr}\,X}|\,.
$$

\subsubsection*{Gauginos and the fermions $\psi_{X'}$}

There are no mixings between the gauginos and the $\psi_{X'}$ fermions at one and two loops, because they are forbidden by charge conjugation. The expression for the one loop gaugino mass is
\begin{equation}\label{eq:mg}
m_{\lambda}=\,\frac{2g^2\tilde{N}_c}{16\pi^2}\,\sum_{c=1}^2\sum_{d=1}^2\sum_{j=1}^4\sum_{k=1}^2\, (U_f^*)_{kc}\,(\tilde U_f^*)_{k,d}\,(U_b)_{jc}\,(U_b^*)_{j,d+2}\,I[\tilde m_j,m_k]\,.
\end{equation}
which is of order $g^2 \mu_\phi$. The one loop computation for the masses of $\psi_{X'}$ is nearly identical to that of $\psi_{{\rm tr}\,X}$, given in (\ref{eq:loop-mG}), since they have the same interactions with the messenger fields. The result is
\begin{equation}
m_{\psi_{X'}} = h^2\mu_\phi-\frac{2h^2\tilde{N}_c}{16\pi^2}\,\sum_{j=1}^4\sum_{k=1}^2\, (U_f^*)_{k1}\,(\tilde U_f^*)_{k1}\,(U_b^*)_{j1}\,(U_b)_{j3}\,\,I[\tilde m_j,m_k]
\end{equation}
The cancellation that occurs in (\ref{eq:loop-mG}) occurs here as well, but leaves over a large remainder, of order $|\gamma \mu_\phi|$.

\end{document}